\documentclass[reprint,nofootinbib,superscriptaddress,floatfix,preprintnumbers,amsmath,amssymb,10pt]{revtex4-1}
\usepackage[titletoc,toc,title]{appendix}
\usepackage[english]{babel}
\usepackage[hidelinks]{hyperref}
\usepackage[normalem]{ulem}
\usepackage{adjustbox}
\usepackage{amsfonts,amsmath,amssymb}
\usepackage{bm}
\usepackage{enumitem}
\usepackage{epsfig}
\usepackage{cancel}
\usepackage{centernot}
\usepackage{color}
\usepackage{comment}
\usepackage{contour}
\usepackage{flushend}
\usepackage{graphics}
\usepackage{graphicx}
\usepackage{mathrsfs}
\usepackage{mdframed}
\usepackage[normalem]{ulem}
\usepackage{pifont}
\usepackage{relsize}
\usepackage{shadowtext}
\usepackage{slashed}
\usepackage{soul}
\usepackage{subfigure}
\usepackage{tcolorbox}
\usepackage{textgreek}
\usepackage{titlesec}
\usepackage{titletoc}
\usepackage{verbatim}
\usepackage{xcolor}
\usepackage{xfrac}
\contourlength{0.2em}
\usepackage{orcidlink}
\usepackage{ragged2e} 

\definecolor{zima_blue}{HTML}{1393C1}
\hypersetup{setpagesize=false,bookmarksnumbered=true,bookmarksopen=true,colorlinks=true,linkcolor=zima_blue,urlcolor=zima_blue,citecolor=zima_blue,linktocpage=false}

\begin{document}

\author{Fatemeh Elahi}
\email{felahi@uni-mainz.de}
\affiliation{PRISMA$^{++}$ Cluster of Excellence $\&$ Mainz Institute for Theoretical Physics, Johannes Gutenberg University, 55099 Mainz, Germany
}
\author{Pedro Schwaller}
\email{pedro.schwaller@uni-mainz.de}
\affiliation{PRISMA$^{++}$ Cluster of Excellence $\&$ Mainz Institute for Theoretical Physics, Johannes Gutenberg University, 55099 Mainz, Germany
}

\title{A Vector-Like Lepton Interpretation of the High-Energy
Nuclear Recoil Candidate in LUX-ZEPLIN
}

\begin{abstract}
The LUX-ZEPLIN (LZ) experiment has recently reported a nuclear-recoil candidate at $E_R=248\pm23_{\rm stat}\pm23_{\rm sys}\,{\rm keV}$ in an exposure of $2.84\,{\rm tonne\,yr}$. At this unusually high recoil energy, the spin-independent (SI) xenon response is strongly suppressed, motivating dark matter interactions with a harder recoil spectrum. We show that elastic spin-dependent (SD) scattering through the nonrelativistic operator $\mathcal O_4=\mathbf S_{\rm DM}\cdot\mathbf S_N$ provides a simple realization of such a spectrum. We consider singlet-doublet Majorana dark matter, for which a diagonal $Z$ coupling generates the required SD interaction. The accompanying Higgs-mediated SI interaction would generically produce too many low-energy recoils, but can be suppressed along a Higgs blind spot while retaining a nonzero $Z$ coupling. We find a region of the Higgs blind spot that simultaneously predicts an $\mathcal O(1)$ event rate in the LZ high-recoil window and reproduces the observed thermal relic abundance. The same SD interaction leads to solar capture, allowing this region to be tested independently with solar-neutrino searches. Current IceCube limits already probe part of the LZ-compatible thermal parameter space, while a viable region remains. Additional xenon exposure and improved solar-neutrino searches can therefore provide complementary tests of this interpretation.
\end{abstract}

\preprint{MITP-26-043}
\maketitle
\section{Introduction}

The LUX-ZEPLIN (LZ) experiment has recently reported a nuclear-recoil candidate with reconstructed recoil energy~\cite{LZ:2026axp}
\begin{equation*}
     E_R = 248 \pm 23_{\rm stat} \pm 23_{\rm sys}\,{\rm keV},
\end{equation*}
in a total exposure of $2.84\,{\rm tonne\,yrs}$. The event lies near the upper end of the recoil-energy range considered in the analysis and corresponds to a maximum local significance of approximately $3.4\sigma$, with a global background-only significance of $2.6\sigma$. While the statistical significance is not sufficient to claim evidence for dark matter, the event provides an interesting target for particle-physics interpretations.

The unusually large recoil energy is particularly interesting. For a xenon target, the corresponding momentum transfer is $q = \sqrt{2m_{\rm Xe}E_R}\simeq 0.25~{\rm GeV},$ placing the event in a regime where the nuclear response plays an important role in determining the shape of the recoil spectrum.

Recent analyses of the LZ event have found that elastic spin-dependent scattering through the nonrelativistic operator
$\mathcal O_4 = \mathbf S_{\rm DM}\cdot\mathbf S_N$ ~\cite{Fan:2010gt,Anand:2013yka,LZ:2026axp}, where $\mathbf{S}$ is the spin operator, 
can provide a substantially better description of the high-energy event than conventional elastic spin-independent scattering, with a local significance of $2.7~\sigma$ for DM masses above $400$~GeV. In the relativistic operator basis considered by LZ, this interaction arises from $\mathcal{L}_{15}$, corresponding to an axial-vector DM current coupled to the axial-vector quark current~\cite{LZ:2026axp}.
This motivates a simple question: can the interaction favored by the high-recoil event arise naturally from a well-defined particle-physics model with a viable thermal history? In this work we show that it can.  We consider the vector-like lepton model of Ref.~\cite{Joglekar:2012vc}, in which the dark matter candidate is a Majorana fermion arising from mixing between electroweak singlet and doublet states\footnote{See also~\cite{Arkani-Hamed:2012dcq}. Related models often go by the name of singlet-doublet DM~\cite{Cohen:2011ec}, see e.g.~\cite{Bhattiprolu:2025beq,Griffith:2026hdi} for some recent studies.}. The doublet component induces a diagonal axial coupling to the $Z$ boson, $ Z_\mu \overline{N}_1\gamma^\mu\gamma^5 N_1,$ which at low energies gives precisely the $\mathcal O_4$ interaction.

We find a region that simultaneously gives a significant thermal relic abundance and an O(1) event rate in the high-recoil LZ window, without overproducing lower-energy recoils constrained by earlier LZ searches~\cite{LZ:2024zvo}. The same coupling to the $Z$ boson also allows solar capture of the DM, whose subsequent annihilation predicts an increased flux of high energy neutrinos, which we confront with limits from IceCube~\cite{IceCube:2025fcu}. We comment on the prospects for probing this scenario with upcoming data from direct detection experiments, neutrino telescopes and in direct searches for the charged co-annihilation partners at LHC and future colliders. 

Our interpretation is qualitatively different from the recently proposed endothermic and exothermic inelastic dark matter interpretations of the LZ event~\cite{Fan:2026kxx,Yin:2026jnn,Rodd:2026tyn,Du:2026guj,DiMauro:2026ldr,Jeesun:2026vzo,McCabe:2026crm,Smirnov:2026aqk,Nomura:2026qyq,Visinelli:2026kgt,Bose:2026ndd,Wang:2026ytg,Das:2026uyy,Langhoff:2026ujr,Chatterjee:2026scv,Ahmed:2026qjg,Dent:2026bji,deLima:2026shq}. In these scenarios, the high recoil energy is generated by inelastic kinematics, whereas here the scattering is elastic and the spectral feature arises from the momentum dependence of the xenon nuclear response. Dark matter absorption, $\chi N\to\nu N$, provides another mechanism for generating a high-energy recoil~\cite{Lou:2026idn}, although the LZ-compatible parameter space is strongly constrained by KamLAND. In our scenario the scattering is instead elastic, a less explored option so far~\cite{Unwin:2026rdp}. The high-energy recoil is therefore a consequence of the elastic spin-dependent nuclear response rather than an inelastic mass threshold. Solar capture provides an independent constraint on dark matter interpretations of the LZ event and has recently been emphasized in the context of inelastic scattering ~\cite{Pospelov:2026ewn,Nguyen:2026lui}. This constraint is also relevant for our scenario: the same elastic SD interaction responsible for the LZ signal captures dark matter in the Sun, providing a complementary probe through solar-neutrino searches.

\section{Model Framework}

We consider an extension of the Standard Model (SM) by vector-like leptons following~\cite{Joglekar:2012vc}, which adds a SM like fourth generation and its mirror copy to the SM. For DM phenomenology, it is sufficient to consider a subset of states, namely a vector-like doublet lepton and a single Majorana neutrino, also known as singlet-doublet DM\footnote{The thermal history and phenomenological constraints of singlet-doublet Majorana dark matter have also been studied recently in Ref.~\cite{Paul:2025spm}}~\cite{Cohen:2011ec}. A parity symmetry under which only the new states are odd guarantees the stability of the lightest  particle and forbids mixing with SM leptons. 
The new fermions consist of a gauge-singlet Majorana state $\nu\sim({\bf 1},{\bf 1},0)$ and a vector-like electroweak doublet $\ell_{L,R}^T=(\nu_{L,R},e_{L,R})$, transforming as $\ell_{L,R}\sim({\bf 1},{\bf 2},-1/2)$. The relevant terms in the Lagrangian are
\begin{equation}
\mathcal L\supset-m_\ell\bar\ell_L\ell_R-\frac12m_0\bar\nu^c\nu-\left(Y_n\bar\ell_L\widetilde H\nu+Y_n'\bar\ell_R \widetilde H\nu^c+\mathrm{h.c.}\right),
\end{equation}
where $\widetilde H=i\sigma_2H^*$ and $H=(0,(v+h)/\sqrt2)^T$ with $v=246$ GeV. After electroweak symmetry breaking, the neutral states mix through
\begin{equation}
\mathcal M_n=
\begin{pmatrix}
0&m_\ell&{Y_nv}/{\sqrt2}\\
m_\ell&0&{Y'_nv}/{\sqrt2}\\
{Y_nv}/{\sqrt2}&{Y'_nv}/{\sqrt2}&m_0
\end{pmatrix},
\end{equation}
in the basis $(\nu_L,\nu_R^c,\nu)^T$. The symmetric mass matrix is diagonalized by $V^T\mathcal M_nV=\mathrm{diag}(M_{N_1},M_{N_2},M_{N_3})$, where $N_1$ is the lightest neutral state and the dark matter candidate. The charged state, denoted by $E$, has mass $M_E\simeq m_\ell$, up to electroweak radiative corrections.

The coupling relevant for the high-energy nuclear recoils is the diagonal $Z$ interaction. Since the singlet does not couple to the $Z$, the coupling of $N_1$ is determined by its doublet components. For a Majorana fermion the diagonal vector current vanishes, leaving
\begin{equation}
\mathcal L \supset\frac{g}{4c_W} g_{ZNN} \,\overline N_1\gamma^\mu\gamma^5N_1Z_\mu,
\qquad
g_{ZNN}=|V_{11}|^2-|V_{12}|^2.
\label{eq:Zcoupling}
\end{equation}
The resulting non-relativistic interaction is predominantly spin dependent (SD). The same mixing also generates a Higgs coupling and hence a spin independent (SI) scattering contribution: 
\begin{equation}
\mathcal L \supset\frac{1}{2} g_{hNN} \,\overline N_1 N_1 h,
\ \ \ 
g_{hNN}=V_{13}(Y_n V_{11} - Y_n' V_{12}).
\label{eq:hcoupling}
\end{equation}
In the limit $Y_nv,Y_n'v\ll|m_0-m_\ell|$, the couplings of the mostly-singlet lightest state can be written as~\cite{Bhattiprolu:2025beq}
\begin{align}
g_{ZNN}&=\frac{v^2}{2}\frac{Y_n^2-Y_n'^2}{m_0^2-m_\ell^2},\\
g_{hNN}&=\frac{v}{\sqrt2}\frac{(Y_n^2+Y_n'^2)m_0+2Y_nY_n'm_\ell}{m_0^2-m_\ell^2}.
\label{eq:hZcouplings}
\end{align}
While $g_{ZNN}$ controls the desired SD signal, $g_{hNN}$ induces SI scattering that would preferentially populate the lower-recoil region. It is therefore important that the two couplings can be varied independently enough to suppress the latter without eliminating the former.

In particular, the Higgs coupling vanishes along a tree-level blind spot, $g_{hNN}=0$~\cite{Cohen:2011ec,Cheung:2012qy,Cheung:2013dua,Calibbi:2015nha,Bhattiprolu:2025beq}. Defining $r=Y_n'/Y_n$, this condition gives
\begin{align}
r^2+2\frac{m_\ell}{m_0}r+1=0,
\ \ \
r_\pm=-\frac{m_\ell}{m_0}\pm\sqrt{\left(\frac{m_\ell}{m_0}\right)^2-1},
\label{eq:blindspot}
\end{align}
which has real solutions for $m_\ell>m_0$. Along these branches the Higgs-mediated SI interaction vanishes at tree level, while $g_{ZNN}$ remains nonzero away from the degenerate point $m_\ell=m_0$, where $Y_n'=-Y_n$ and the $Z$ coupling vanishes as well.\footnote{The blind-spot condition in Eq.~\eqref{eq:blindspot} is defined at tree level. Electroweak corrections are known to shift the tree-level blind spot in singlet-doublet dark matter rather than remove it~\cite{Han:2018gej}. For parameters comparable to those considered here, the resulting displacement of the blind-spot condition was found to be at the percent level. We therefore expect the tree-level blind-spot region identified here to persist after radiative corrections, with a perturbative shift of its location. A dedicated NLO calculation for our parameter space is beyond the scope of this work.} Thus the model admits a genuine separation between the two direct-detection channels: the SI interaction can be strongly suppressed while retaining a phenomenologically relevant SD coupling.

The mass splitting $\Delta m\equiv m_\ell-m_0$ simultaneously controls the thermal history. For small $\Delta m$, the charged and heavier neutral states remain in equilibrium during freeze-out, and coannihilation with these electroweak partners efficiently depletes the dark matter abundance. The same mass and mixing parameters therefore determine the relic abundance, the $Z$-mediated SD signal, and the Higgs-mediated SI contribution. This interplay is central to the parameter region considered below.

\section{High-Energy Nuclear Recoils}

\begin{figure*}[t]
    \centering
    \includegraphics[width=0.9\linewidth]{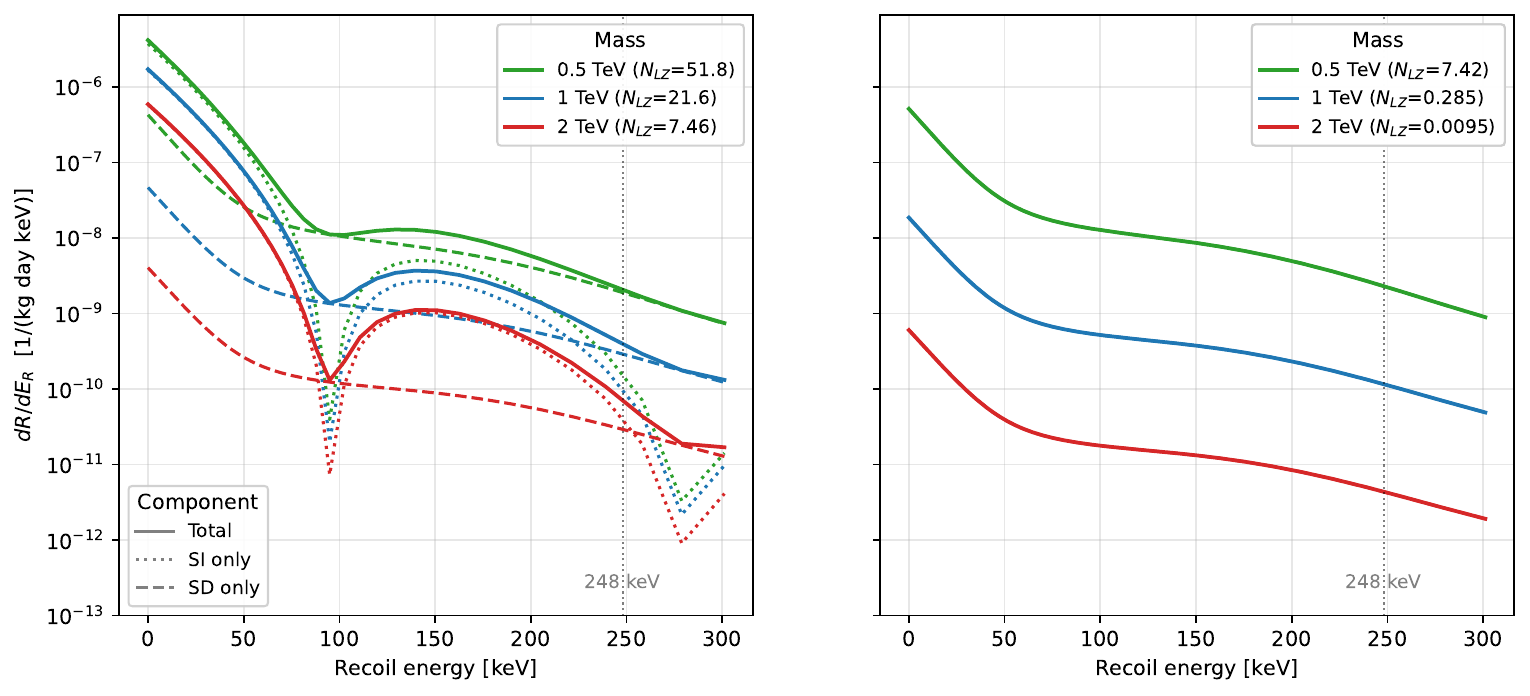}
   \caption{Predicted xenon recoil spectra illustrating the role of the Higgs blind spot. 
\textit{Left:} Without the Higgs blind spot ($Y_n'=0$), Higgs-mediated SI scattering dominates at low recoil energies, leading to a large event yield; for $M_{\rm DM}=0.5~{\rm TeV}$, for example, the predicted number of events is $N_{\rm LZ}=46.7$. 
\textit{Right:} Choosing $Y_n'$ on the blind-spot branch such that $g_{hN_1N_1}=0$ removes the tree-level SI contribution while retaining the $Z$-mediated SD interaction. For the same $0.5~{\rm TeV}$ point, the event yield is reduced to $N_{\rm LZ}=1.27$, while the recoil spectrum remains appreciable at $E_R\simeq248~{\rm keV}$ (vertical dotted line). 
 In both panels $Y_n=0.1$ and $(m_\ell - m_0)/m_0=0.01$.}
    \label{fig:recoil_spectra}
\end{figure*}

The LZ candidate at $E_R\simeq248~{\rm keV}$ corresponds to a momentum transfer $q=\sqrt{2m_{\rm Xe}E_R}\simeq245~{\rm MeV}$~\cite{LZ:2026axp}. At such large momentum transfer, the wavelength of the exchanged momentum becomes comparable to nuclear length scales, and the scattering no longer probes the nucleus as a fully coherent object. Interference between amplitudes from different parts of the nucleus therefore becomes important. For the SI response of xenon, the LZ candidate happens to lie close to a destructive-interference minimum of the Helm form factor: $F_{\rm SI}^2(q\simeq245~{\rm MeV})\ll1$~\cite{Helm:1956zz,Lewin:1995rx}. An SI interaction normalized to produce an event near $250~{\rm keV}$ therefore generically predicts a much larger population at lower recoil energies, where the form-factor suppression is absent. The observation of a high-energy event without a corresponding low-energy excess disfavors such spectra.

The situation is qualitatively different for spin-dependent (SD) scattering. The spin response of the xenon isotopes with nonzero nuclear spin remains appreciable at $q\simeq245~{\rm MeV}$, making an SD interaction particularly well suited to populate the high-recoil region. The singlet-doublet model provides precisely such an interaction: the diagonal $Z$ coupling in Eq.~\eqref{eq:Zcoupling} generates predominantly the non-relativistic operator $\mathcal O_4=\mathbf S_{\rm DM}\cdot\mathbf S_N$.

Eventhough $O_4$ solves the nuclear-response problem, the UV model reintroduces SI through Higgs exchange, which repopulates the low-recoil region. Therefore the blind spot is necessary for the LZ interpretation.  This is illustrated in the left panel of Fig.~\ref{fig:recoil_spectra}: without suppressing the Higgs coupling, parameter points capable of producing an event in the high-recoil region predict many more events at lower energies. The Higgs blind spot of Eq.~\eqref{eq:blindspot} removes this contribution while retaining a nonzero diagonal $Z$ coupling. The resulting spectrum, shown in the right panel of Fig.~\ref{fig:recoil_spectra}, is then dominated by SD scattering and remains comparatively flat over the recoil range relevant for the LZ candidate. The spectra are obtained using \texttt{micrOMEGAs}~\cite{Belanger:2010pz}. 

We determine the LZ signal from the full recoil spectrum, including the xenon isotope abundances, nuclear response functions, and detector acceptance, rather than from the nucleon cross section alone. Viable points are required to yield an $\mathcal O(1)$ event rate in the high-recoil region without producing an excluded population at lower recoil energies. As we show below, this requirement can be satisfied simultaneously with the thermal relic abundance.

\section{Relic Density and the LZ-Compatible Parameter Region}

We now ask whether the parameter region producing the high-recoil LZ signal can simultaneously account for the thermal dark matter abundance. We fix the relative mass splitting to $\Delta m/m_0=0.01$ and impose the Higgs blind-spot condition of Eq.~\eqref{eq:blindspot}. Since this condition admits two solutions, $r_\pm=Y_n'/Y_n$, we scan both branches separately.
The relic abundance is calculated with \texttt{micrOMEGAs} \cite{Belanger:2010pz}. Sommerfeld enhancement is almost negligible for pure Higgsino dark matter~\cite{Cirelli:2007xd,Hryczuk:2010zi}, and mixing with a singlet should further reduce the effect. We therefore do not include a Sommerfeld correction factor in our analysis, but allow a 10\% uncertainty band for the relic density prediction, which should be sufficient to cover this as well as other uncertainties.

\begin{figure*}[t]
    \centering
    \includegraphics[width=0.95\linewidth]{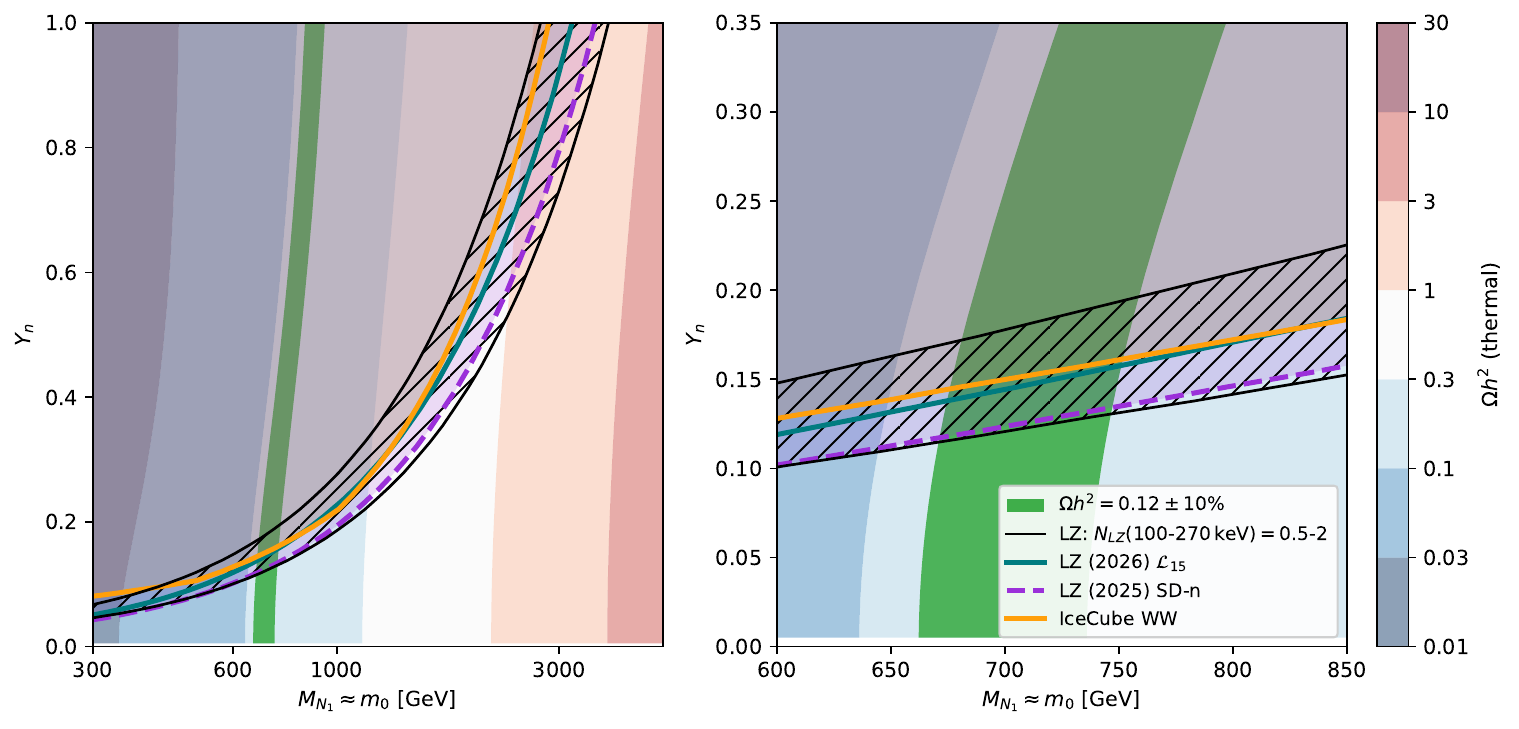}
    \caption{ Relic abundance, LZ event rate, and direct- and indirect-detection constraints along the $r_+$ Higgs-blind-spot branch of Eq.~\eqref{eq:blindspot}, for a fixed mass splitting $\Delta m/m_0=0.01$. The color scale shows the thermal relic abundance $\Omega_{\rm DM}h^2$, with the green band indicating $\Omega_{\rm DM}h^2=0.12\pm10\%$. The hatched region corresponds to $0.5<N_{\rm LZ}<2$ events in the $100- 270~{\rm keV}$ recoil window. The cyan curve shows the LZ constraint on the $L_{15}$ interaction~\cite{LZ:2026axp}, while the purple dashed curve shows the LZ low-energy SD-neutron limit~\cite{LZ:2024zvo}. The orange curve denotes the IceCube solar-neutrino constraint for annihilation into $W^+W^-$~\cite{IceCube:2025fcu}. For each constraint, the region above the corresponding curve is excluded. The right panel zooms in on the region where the thermal relic-density and LZ event-rate bands overlap, showing the remaining parameter space consistent with the current LZ and IceCube constraints. }
    \label{fig:relicandLZ}
\end{figure*}

The result is shown in Fig.~\ref{fig:relicandLZ}. The color scale gives the thermal relic abundance in the $(M_{N_1},Y_n)$ plane, with the green band denoting $\Omega h^2=0.12\pm10\%$. The hatched region corresponds to $0.5<N_{\rm LZ}<2$ events in the $100$--$270~{\rm keV}$ recoil window. We only show results for the $r_+$ blind spot branch, since the $r_-$ branch looks almost identical.

The relic-density and LZ requirements select parametrically different regions. For the small mass splitting considered here, the relic abundance is largely controlled by coannihilation with the nearly degenerate charged and heavier neutral states, while the LZ rate is controlled by the doublet admixture that determines the diagonal $Z$ coupling. Nevertheless, the two regions overlap for sub-TeV dark matter masses and $Y_n=\mathcal O(0.1)$. Importantly, this overlap is present on both blind-spot branches. Thus suppressing the unwanted Higgs-mediated SI interaction does not remove either the SD signal required by LZ or the thermal dark matter solution.

At larger masses the relic abundance increases rapidly as electroweak annihilation and coannihilation become less efficient. The overlap in Fig.~\ref{fig:relicandLZ} therefore identifies the region in which the three requirements central to our interpretation are simultaneously satisfied: a suppressed SI interaction, an $\mathcal O(1)$ high-recoil LZ rate, and the observed thermal relic abundance. We additionally impose the LZ constraint on the relativistic axial--axial operator $L_{15}$, which matches onto the $\mathcal O_4$ interaction relevant here. We adopt the $L_{15}$ normalization that the dimension-six interaction is suppressed by $v^{-2}$, where $v=246~{\rm GeV}$, and match the interaction in our model onto the corresponding coefficient $d_{15}$. The resulting constraint is shown by the cyan curve in Fig.~\ref{fig:relicandLZ}.

Of course non-standard cosmologies can easily modify the relic abundance without affecting the direct detection rate. Since most of the parameter space is hampered by DM overproduction, let us briefly mention two possible escape routes. First, if the universe is reheated to temperatures significantly below the DM mass, which is a perfectly viable scenario, the DM abundance will be suppressed, making the above TeV region viable. A second possibility is entropy injection into the SM plasma after DM freeze-out, such as from a strongly supercooled phase transition~\cite{Huang:2026qdc}, which can dilute an initially over-abundant DM candidate. Such a scenario has the additional benefit of predicting a stochastic gravitational wave signal which can be probed experimentally.

\section{Constraints and Prospects}

Dark matter scattering in the Sun provides an important complementary constraint on the parameter space relevant for the LZ signal, and was recently used in Ref.~\cite{Pospelov:2026ewn} to strongly constrain its interpretation in terms of inelastic Higgsino dark matter. DM particles passing through the Sun are accelerated by its gravitational potential and may lose sufficient kinetic energy in scattering with solar nuclei to become gravitationally bound. Subsequent scatterings cause the
captured population to accumulate and thermalize near the solar core, where dark matter annihilation can produce high-energy neutrinos observable at neutrino telescopes such as IceCube~\cite{IceCube:2025fcu}. 

In our model, capture proceeds through the same elastic SD interaction that generates the LZ signal. Since the Sun is composed predominantly of hydrogen, the capture rate is particularly sensitive to the DM--proton SD cross section\footnote{This differs from the inelastic scenario of Ref.~\cite{Pospelov:2026ewn}, where the large velocities attained in the Sun allow scattering above the endothermic threshold and make heavy solar elements particularly important.}~\cite{Griest:1986yu,Silk:1985ax,Krauss:1985ks,Gould:1987ju,Peter:2009mk,Zentner:2009is,Busoni:2013kaa,Garani:2017jcj,Busoni:2017mhe}. The number of captured particles is determined by the competition between capture and annihilation: $\dot N_{\rm DM}=C_\odot-C_A N_{\rm DM}^2$. The capture rate $C_\odot$ is controlled primarily by the local dark matter flux and the scattering cross section:  $C_\odot\propto(\rho_{\rm DM}/m_{\rm DM})\sigma_{\rm SD}^p$. Annihilation competes with this accumulation at a rate $\Gamma_{\rm ann}=C_A N_{\rm DM}^2/2$, where $C_A\simeq\langle\sigma v\rangle_0/V_{\rm eff}$  is determined by the present-day annihilation cross section and the volume occupied by the captured dark matter in the solar core. Initially, capture dominates and dark matter accumulates in the Sun. As the population grows, however, the annihilation rate increases as $N_{\rm DM}^2$, eventually balancing the capture rate. Defining the equilibration time $\tau_\odot=(C_\odot C_A)^{-1/2}$, the annihilation rate is
$$ \Gamma_{\rm ann}=\frac{C_\odot}{2}\tanh^2\!\left(\frac{t_\odot}{\tau_\odot}\right). $$
For the parameter region relevant here, we find $\tau_\odot$, well below the solar age $t_\odot\simeq4.6~{\rm Gyr}$. Capture and annihilation have therefore reached equilibrium, and $\Gamma_{\rm ann}\simeq C_\odot/2$. Physically, this means that the Sun has had enough time for annihilations to keep pace with capture: once equilibrium is reached, increasing or decreasing $\langle\sigma v\rangle_0$ does not appreciably change the neutrino signal, which is instead set primarily by the SD scattering cross section that also controls the LZ signal. 
For the parameter region relevant here, we find $\tau_\odot$, well below the solar age $t_\odot\simeq4.6~{\rm Gyr}$. Capture and annihilation have therefore reached equilibrium, and $\Gamma_{\rm ann}\simeq C_\odot/2$. Physically, this means that the Sun has had enough time for annihilations to keep pace with capture: once equilibrium is reached, increasing or decreasing $\langle\sigma v\rangle_0$ does not appreciably change the neutrino signal, which is instead set primarily by the SD scattering cross section that also controls the LZ signal. 

The recent ten-year IceCube solar analysis constrains SD scattering at the level of $\sigma_{\rm SD}^p\sim10^{-41}~{\rm cm^2}$ for annihilation into $W^+W^-$ in the mass range relevant here~\cite{IceCube:2025fcu}. As shown by the orange curve in Fig.~\ref{fig:relicandLZ}, this constraint already excludes part of the parameter space compatible with both the LZ
event and the thermal relic abundance, while a viable region remains. Solar-neutrino searches therefore provide an immediate and complementary test of the LZ interpretation~\cite{Bell:2021pyy}.

The small present-day annihilation rate also illustrates an important distinction from thermal freeze-out. In the compressed region considered here, the relic abundance is strongly affected by coannihilation with the heavier electroweak states, which are absent from the Galactic dark matter population today. Consequently, the present-day $N_1N_1$ annihilation rate is about an order of magnitude below the canonical thermal value, suppressing conventional halo indirect-detection signals.

The most direct test remains additional data from xenon based direct detection experiments. Our scenario predicts an appreciable amount of low energy recoil events, which we confront with SD limits from the LZ low energy recoil search~\cite{LZ:2024zvo}. Similar to the IceCube constraints, this excludes the upper end of our signal region, shown by the purple contour in Fig.~\ref{fig:relicandLZ}. It is worth noting that our best fit region is sensitive to the recoil energy window chosen. Here we chose the region between 100~keV and 270~keV, across which the recoil spectrum is relatively flat, while the reconstruction efficiency stays above 50\%~\cite{LZ:2026axp}. A more narrow window would increase the tension with LZs SD limit and IceCube data, while extending the window down to 55~keV recoil energy (the upper limit of the previous search) would relax the tension. 

If the LZ candidate originates from $N_1$, future data should populate the same SD recoil spectrum rather than reveal the steep low-energy spectrum characteristic of an unsuppressed SI interaction. Together with the near-term sensitivity of solar-neutrino searches, this provides two complementary tests of the blind-spot interpretation. The solar annihilation constraint could however be further weakened by additional annihilation channels into invisible dark-sector states, which reduce the branching fraction into neutrino-producing SM final states without altering the Z-mediated scattering responsible for LZ~\cite{Bell:2021pyy}. Since freeze-out in the relevant region is dominated by coannihilation, such an extension need not qualitatively modify the relic-density mechanism.

Let us for completeness discuss electroweak precision constraints. While extensions of the SM with chiral fermions are strongly constrained due to non-decoupling, in the limit where the vector-like masses dominate the contributions to the $S$ and $T$ parameters are suppressed. Since the best fit to the LZ event requires masses above 400~GeV, we are safely in this regime already~\cite{Joglekar:2012vc}. Contributions to Higgs precision observables are also under control in this case, in particular since the charged lepton Yukawas, which can significantly enhance the Higgs to $\gamma\gamma$ and Z$\gamma$ decays, are set to zero for now. A more careful study of possible indirect probes of this scenario with future LHC data might nevertheless be useful.

\section{Collider probes}
\begin{figure}[t]
    \centering
    \includegraphics[width=\linewidth]{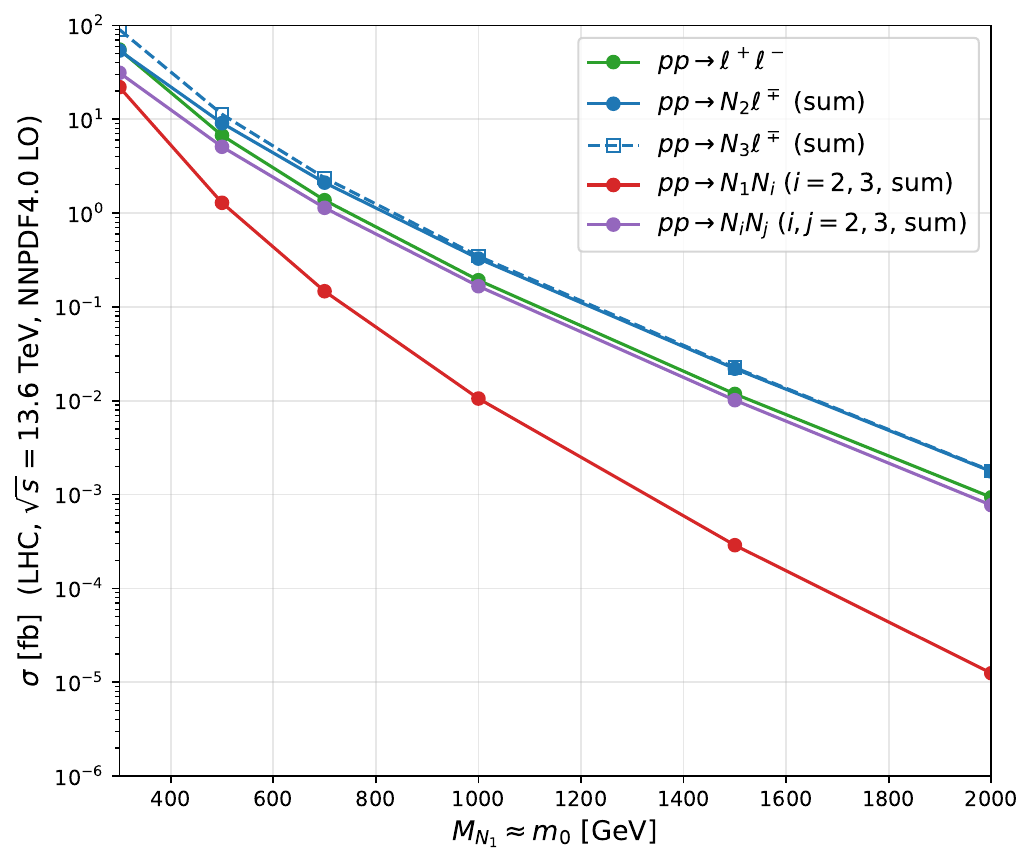}
    \caption{Cross sections for pair production of heavy lepton pairs at the LHC with 13.6~TeV centre-of-mass energy. Shown are the rates for charged lepton pairs (green), neutral lepton pairs $N_iN_j$ and charged neutral pairs $\ell^\pm N_i$. }
    \label{fig:LHC}
\end{figure}

At the LHC, the new leptons are pair produced via exchange of $W/Z$ bosons and photons. Direct production of $N_1$ pairs is suppressed and also irrelevant, since detecting these events requires additional jets from initial state radiation. Instead pairs of charged or charged and neutral heavy leptons can be produced copiously, for masses up to the TeV scale. Cross sections for the LHC at 13.6~TeV centre-of-mass energy were obtained using CalcHEP~\cite{Belyaev:2012qa} with NNPDF4.0, and are shown in Fig.~\ref{fig:LHC}. 

The mass splitting between $N_1$ and the heavier leptons is between a few and tens of GeV. It is sufficiently large that the heavier states will promptly decay to $N_1$ via virtual $W/Z$ emission. The smoking gun signature is soft lepton events with large amounts of missing energy, with current limits reaching masses up to 300~GeV~\cite{ATLAS:2019lng}. While future LHC searches for electroweak new physics could reach up to TeV scale masses~\cite{Liu:2020muv}, the prospects are somewhat more modest for compressed spectra such as the one considered here~\cite{Schwaller:2013baa}. 

Among the future colliders currently under consideration, a 100~TeV hadron collider like FCC-hh or a muon collider with centre-of-mass energy of 3~TeV or more offer the best chances for probing this dark matter scenario. The FCC-hh can even probe the nightmare scenario of highly degenerate states by combining mono-jet~\cite{Low:2014cba} and disappearing track searches~\cite{Mahbubani:2017gjh}. The additional soft leptons expected here should boost the sensitivity into the multi-TeV region. Similarly, the clean environment of a muon collider~\cite{Accettura:2023ked} should guarantee discoveries up to half of its centre-of-mass energy~\cite{Capdevilla:2024bwt}.

\section{Conclusions}

The nuclear-recoil candidate reported by LZ is particularly intriguing because its energy, $E_R\simeq248~{\rm keV}$, corresponds to a momentum transfer where the SI response of xenon is strongly suppressed. If the event originates from dark matter, interactions whose nuclear response remains appreciable at large momentum transfer are therefore particularly well motivated. We have shown that elastic spin-dependent scattering provides a simple realization of this possibility.

In the singlet-doublet DM model, the required interaction arises from the diagonal $Z$ coupling and is dominated by
$\mathcal O_4=\mathbf S_{\rm DM} \cdot\mathbf S_N$. The same singlet-doublet mixing generically induces Higgs-mediated SI scattering, which would instead produce a large population of lower-energy recoils. The two Yukawa couplings of the model allow this contribution to be suppressed at a Higgs blind spot while retaining a nonzero SD interaction. We find that, on both branches of the blind spot, the region producing an $\mathcal O(1)$ event rate in the LZ high-recoil window overlaps with the
thermal relic-density band. The relic abundance in this region is set largely by coannihilation with the nearly degenerate electroweak states. 

The LZ collaboration performed fits to 20 relativistic DM-nucleon interaction Lagrangians~\cite{LZ:2026axp}. Our scenario corresponds to ${\mathcal L}_{15}$, which reduces to $\mathcal O_4$ in the non-relativistic limit. Here we showed how to obtain this effective interaction from a renormalizable and gauge invariant model. We also find that a UV completion normally generates additional interactions, which may spoil the simplest effective-operator interpretation but also lead to correlated observables. 

An important consequence is that the same SD interaction responsible for the LZ signal also leads to efficient capture in the Sun. As shown in Fig.~\ref{fig:relicandLZ}, current IceCube solar-neutrino limits already probe part of the parameter space where the LZ event rate and thermal relic abundance overlap, while a viable region remains. Solar-neutrino searches therefore provide a direct and complementary test of this interpretation. We also find that the interpretation is slightly in tension with LZs own limits on SD DM scattering. 
Additional data from the LZ and XENONnT experiments should therefore quickly reveal whether the observed event is due to DM with elastic SD interactions. Further events should follow the characteristic SD recoil spectrum rather than the steep low-energy spectrum expected from an unsuppressed SI interaction, but with a sizable number of events expected also at low recoil energies. 

Taken together, these observations provide a simple and testable interpretation of the LZ candidate. If the high-recoil event persists with additional exposure, its recoil spectrum and the associated solar-neutrino signal can test not only a dark matter origin, but also the spin structure of the underlying interaction. The LHC and future colliders would allow further scrutiny of these scenarios. 

Let us conclude with some speculation about the UV embedding of this model. Vector-like leptons appear in many extensions of the SM such as composite Higgs or models of flavour, and can often be studied in isolation~\cite{Falkowski:2013jya}. One interesting scenario is models with gauged lepton number~\cite{FileviezPerez:2010gw,Schwaller:2013hqa}, where they are essential for the cancellation of gauge anomalies, and the gauge symmetry, even after spontaneous breaking, guarantees the stability of dark matter. This additional gauge symmetry however forbids Majorana masses, such that the DM candidate would become a Dirac fermion, with potentially unsuppressed spin-independent direct detection rates. Therefore in this scenario Majorana masses should be generated when lepton number is spontaneously broken. The additional leptons in our model also appear as incomplete SU(5) multiplets, so they will spoil grand unification. This can easily be ameliorated by adding vector-like quarks at a few TeV to complete the families. If the LZ event is really our first glimpse at the dark sector, there will be many interesting avenues to explore. 

\textit{\textbf{Acknowledgments.}} We thank M.~Hager, G.~Laverda and C.E.M.~Wagner for useful discussions, and K.~Agashe for discussions regarding the blind spots. We acknowledge the use of Claude (Anthropic) as a coding assistant in the validation of the micrOMEGAs/CalcHEP implementation and for performing parameter scans. Claude (Anthropic) and ChatGPT (OpenAI) were further used to improve the wording in the manuscript. The authors acknowledge support by the Cluster of Excellence ``PRISMA$^{++}$'' funded by the German Research Foundation (DFG) within the German Excellence Strategy (Project No. 390831469).

\bibliographystyle{JHEP}
\bibliography{refs}
\end{document}